\documentclass[12pt]{article}
\begin{document}
\title{ARE WE OBSERVING VIOLATIONS OF LORENTZ SYMMETRY?}
\author{B.G. Sidharth\\
International Institute for Applicable Mathematics \& Information Sciences\\
Hyderabad (India) \& Udine (Italy)\\
B.M. Birla Science Centre, Adarsh Nagar, Hyderabad - 500 063 (India)}
\date{}
\maketitle
\begin{abstract}
Recent observations of ultra high energy cosmic rays and gamma rays suggest that there are small violations of Lorentz symmetry. If there were no such violations, then the GZK cut off would hold and cosmic rays with energy $\sim 10^{20}eV$ or higher would not be reaching the earth. However some such events seem to have been observed. This has lead to phenomenological models in which there is a small violation of the Lorentz symmetry or the velocity of light. However recent Quantum Gravity and String Theory approaches which no longer consider a differentiable spacetime manifold already predict such violations. Similarly there are other theoretical reasons which also point to this. We briefly discuss the various possibilities.
\end{abstract}
\section{The Fuzzy Spacetime Approach}
In a recent communication Pavlopoulos \cite{pav} has suggested that we may already be observing a violation of Lorentz symmetry due to the observed time lags of cosmic gamma rays of different energies. In fact it has been suspected that this could indeed be the case from an observation of ultra high energy cosmic rays. In this case, given Lorentz symmetry there is the GZK cut off such that particles above an energy of $10^{20}eV$ would not be able to travel cosmological distances and reach the earth (Cf. ref.\cite{uof,ijtp} for details). However, it is suspected that some twenty contra events have already been detected, and phenomenological models of Lorentz symmetry violation have been constructed by Glashow, Coleman and others \cite{gon,cole,jacob,olin,car,nag}. The essential point here is that the energy momentum relativistic formula is modified leading to a dispersive effect.\\
We would like to point out that apart from observation based models such a result follows from a fundamental point of view in modern approaches in which the differentiable Minkowski spacetime is replaced by one which is fuzzy or noncommutative owing to a fundamental minimum length $l$ being introduced. This is the case in Quantum Super String Theory and Quantum Gravity approaches (Cf.ref.\cite{uof}). Based on these considerations, we can deduce from theory that the usual energy momentum formula is replaced by $(c = 1 = \hbar)$
\begin{equation}
E^2 = m^2 + p^2 + \alpha l^2 p^4\label{e1}
\end{equation}
where $\alpha$ is a dimensionless constant of order unity. 
This leads to a modification of the Dirac and Klein-Gordon equations at ultra high energies (Cf.ref.\cite{uof,ijtp,ultra}). With this, it has been shown by the author that in the scattering of radiation, instead of the usual Compton formula we have
\begin{equation}
k = \frac{mk_0 + \alpha \frac{l^2}{2}[Q^2 + 2mQ]^2}{[m + k_0 (1 - cos \Theta )]}\label{e2}
\end{equation}
where we use natural units $c = \hbar = 1, m$ is the mass of the elementary particle causing the scattering, $\vec{k} , \vec{k}_0$ are the initial and final momentum vectors respectively and $Q = k_0 - k,$ and $\Theta$ is the angle between the incident and scattered rays. Equation (\ref{e2}) shows that $k = k_0 + \epsilon$, where $\epsilon$ is a positive quantity less than or equal to $\sim l^2$, $l$ being the fundamental length. It must be remembered that in these units $k$ represents the frequency. The above can be written in more conventional form as
\begin{equation}
h \nu = h\nu_0 [1 + 0(l^2)]\label{e3}
\end{equation}
Equation (\ref{e3}) effectively means that due to the Lorentz symmetry violation in (\ref{e1}), the frequency is increased or, the speed of propagation of a given frequency is increased. As noted such models in a purely phenomenological context have been considered by Glashow, Coleman, Carroll and others. In any case what this means in an observational context is that higher frequency gamma rays should reach us earlier than lower frequency ones in the same burst. As Pavlopoulos reports (Cf.ref.\cite{pav}) this indeed seems to be the case.\\
Subject to further tests and confirmation, for example by NASA's GLAST satellite to be launched shortly \cite{glast}, spacetime at a micro or ultra high energy level is not a smooth manifold, brought out by this manifestation in for example (\ref{e1}).
\section{Massive Photons}
Recently the author deduced from theory that the photon has a small mass of $\sim 10^{-65}gms$, this being consistent with the velocity of light and also with experimental limits on the photon mass as will be discussed below \cite{bgfpl,bgsmp}. Interestingly from purely thermodynamic reasoning Landsberg had shown that the above mass is the minimum allowable mass in the universe \cite{land}.  More recently from a completely unrelated point of view, it was shown by the author that this is the minimum allowable mass in the universe in a model where zero point oscillators at the Planck scale provide the underpinning for the universe \cite{bgsfpl,uof}.\\
The derivation of the photon mass uses the fact that for the Langevin equation in the limiting case of low viscous resistance, a particle behaves like a Newtonian particle moving in the absence of any external forces with a uniform velocity given by
$$\langle v^2 \rangle = \frac{kT}{m}$$
If in the above relation we use extreme values of the minimum thermodynamic temperature of the universe and the minimum mass, we recover the velocity of light. To be more specific we use the Beckenstein temperature given by
$$T = \frac{\hbar c^3}{8\pi kMG}$$
with $M \sim 10^{55}gms$, the known mass of the universe. This gives us the value
$$T \sim 10^{-28}K.$$
For the minimum mass we use a result due to Landsberg (Cf.ref.\cite{land}), and also the same result independently obtainable as noted above from a model of Planck oscillators, viz., 
$$m \sim 10^{-65}gms$$
It ust be stressed that though the above equation for the average velocity square resembles the root mean square temperature equation of thermodynamics, it is in fact different. Moreover, this velocity would be maintained for the age of the universe and would be Lorentz invariant \cite{evans} (Cf.ref.\cite{bgfpl} for details).\\
Let us briefly consider some of the consequences of the photon mass and also look for experimental verification, apart from consistency with theory.\\
It may be remarked that the mass for the photon has been proposed in the past, though from phenomenological considerations \cite{evans,bart}. Indeed it is remarkable that exactly the above mass was indicated from experimental observation (Cf.ref.\cite{vig} and references there in), and has been attributed to a vacuum induced dissipative mechanism.\\
With a non zero photon mass we would have, for radiation
\begin{equation}
E = h \nu = m_\gamma c^2 [1 - v^2_\gamma /c^2]^{-1/2}\label{ea1}
\end{equation}
From (\ref{ea1}) one would have a dispersive group velocity for waves of frequency $\nu$ given by (Cf. also ref.\cite{vig})
\begin{equation}
v_\gamma = c \left[1 - \frac{m^2_\gamma c^4}{h^2 \nu^2}\right]^{1/2}\label{ea2}
\end{equation}
We would like to point out that (\ref{ea2}) indicates that higher frequency radiation has a velocity greater than lower frequency radiation. This is a very subtle and minute effect and is best tested in for example, the observation of high energy gamma rays, which we receive from deep outer space. It is quite remarkable that as pointed out, we may already have witnessed this effect- higher frequency components of gamma rays in cosmic rays do indeed seem to reach earlier than their lower frequency counterparts \cite{pav}. The GLAST satellite of NASA to be launched in 2006/2007 may be able to throw more light on these high energy Gamma rays.\\
This apart, a finite photon mass would imply a slight modification of the Coulomb interaction, which would go over into a Yukawwa type potential, this given by, (in natural units $\hbar = 1 = c$.)
\begin{equation}
V (r) = e^{-\mu r}/r\label{ea3}
\end{equation}
where $\mu$ is the mass in these units. As can be seen from (\ref{ea3}), the potential $V$ has a finite range. However this range is $\sim \frac{1}{\mu}$ which is $\sim 10^{28}cms$, as can be easily calculated. This range is in fact the radius of the universe! Nevertheless this cut off does imply that there will not be any infra red divergences \cite{it}.\\
The range of the Yukawa type Coulomb potential being of the order of the radius of the universe, we would expect that the modification would be miniscule. Nevertheless from a strictly mathematical point of view, the photon mass converts the otherwise long range Coulomb potential into a short range Yukawa potential with the consequence that several otherwise strictly divergent integrals become convergent. It then becomes possible to use techniques suitable for short range potentials (Cf. for example \cite{mott,joa}). The rather laborious modifications required for handling the Coulomb potential in scattering theory (Cf. for example \cite{bgs1,bgs2}) can be eased. For instance there is an interesting recurrence relation for the large $l$ phase shifts of the Yukawa potential \cite{bgs3} viz.,
$$\delta_{l+1}/\delta_l \approx 1 - (\mu / K)$$
valid for a large range of energies $K$. This relation can now be used, though as $\mu$ is very small, it shows that the convergence of the phase shifts with respect to $l$ is extremely slow.\\
Another result of the non zero mass of the photon is that in addition to the two traverse polarizations of light, there will be a longitudinal component also (Cf. for example ref.\cite{deser}). It must be remembered that all these effects are small and consistent with the size of the universe. Nevertheless there are experimental tests, in addition to those mentioned above, which are doable. It is well known that for a massive vector field interacting with a magnetic dipole of moment ${\bf M}$, for example the earth itself, we would have with the usual notation (Cf.ref.\cite{it})
$${\bf A}(x) = \frac{\imath}{2} \int \frac{d^3 k}{(2\pi )^3}{\bf M \times k} \frac{e^{\imath k, x}}{{\bf k}^2 + \mu^2} = - {\bf M \times \nabla} \left(\frac{e^{-\mu r}}{8\pi r}\right)$$
\begin{equation}
{\bf B} = \frac{e^{-\mu r}}{8 \pi r^3} | {\bf M}| \left\{\left[ \hat{r} (\hat{r} \cdot \hat{z}) - \frac{1}{3} \hat{z}\right] (\mu^2 r^2 + 3\mu r + 3) - \frac{2}{3} \hat{z} \mu^2 r^2\right\}\label{ea4}
\end{equation}
Considerations like this have yielded in the past an upper limit for the photon mass, for instance $10^{-48}gms$ and $10^{-57}gms$. Nevertheless (\ref{ea4}) can be used for a precise determination of the photon mass. It may be mentioned here that contrary to popular belief, there is no experimental evidence to indicate that the photon mass is zero! (Cf. discussion in ref.\cite{evans}).\\
Finally, it is interesting to observe that the above value for the photon mass was also obtained by Terazawa \cite{tera}, using the Dirac Large Number Hypothesis, something which is in fact a consequence of the Planck oscillator approach alluded to (Cf.ref.\cite{uof}).
\section{The Finsler Spacetime Approach}
In this approach not only the violation of GZK cut off in ultra high energy cosmic rays but also the anisotropy as indicated by data from COBE is taken into account. Then we have a Finsler metric \cite{bg1,bg2,bg3}, which in two dimensions can be written as
\begin{equation}
x' = e^{r\alpha} \quad L(x), tan \, \hbar \alpha = \frac{u}{c}\label{eb1}
\end{equation}
$r$ being the anisotropy factor and $L$ stands for the usual Lorentz transformation, which is not evident in (\ref{eb1}) because there is only one space dimension. From observation it appears that
$$r \geq 10^{-10}$$
Finally the Finslerian metric in three dimensions is given by
\begin{equation}
ds^2 = \left[\frac{(dx^0 - dx^1)^2}{(dx^0)^2 - (dx^1)^2 - (dx^2)^2 - (dx^3)^2}\right]^r \cdot \left[(dx^0)^2 - (dx^1)^2 - (dx^2)^2 - (dx^3)^2\right]\label{eb2}
\end{equation}
It can be seen from (\ref{eb2}) that there is a prefered direction like the $x^1$ axis in this special choice. The metric in (\ref{eb2}) leads to a modified energy momentum formula which is given by
\begin{equation}
\left[\frac{(E/c - \vec{p}\cdot \vec{\nu})^2}{E^2/c^2 - p^2}\right]^{-r} (E^2/c^2 - p^2) = m^2 c^2(1-r)^{1-r} (1+r)^{1+r} (\nu^2 = 1)\label{eb3}
\end{equation}
In (\ref{eb3}) the anisotropy direction is given by $\vec{\nu}$. As $r$ is small, (\ref{eb3}) simplifies to a form similar to (\ref{e1}), with suitable approximations.
\section{Discussion}
We would like to point out that a Lorentz symmetry violation would also imply a violation of the CPT invariance, though this could be expected only from high energies, the effect itself being small \cite{collidy}. Indeed the modified Dirac equation (Cf.ref.\cite{ijtp,ultra}) throws up a Lagrangian with a parity violating term. Specifically, this term in the Dirac Hamiltonian is
$$\gamma^5 l p^2,$$
which is clearly CPT violating.\\
 It must also be remarked that given a fuzzy spacetime or equivalently a noncommutative geometry, we can deduce the photon mass. It has already been pointed out that such noncommutativity of coordinates leads to a term in gauge theory which is similar to the symmetry breaking Higgs field term \cite{uof}. It is this term which in the case of $U(1)$ electromagnetic field gives $\sim l^2$. This is $\sim 10^{-66}gms$, which as argued is the photon mass. 

\end{document}